\documentclass[aps,prl,onecolumn,showpacs,groupedaddress]{revtex4-1}
\usepackage[centertags]{amsmath}
\usepackage{graphicx}
\usepackage{amsfonts}
\usepackage{epstopdf}
\usepackage{ulem}
\usepackage{color}

\usepackage{float}	
\usepackage{amsmath}	
\usepackage{amssymb}
\usepackage{color}

\begin{document}


\title{Initial Wave Breaking Dynamics of Peregrine-Type Rogue Waves: A Numerical and Experimental Study}

\author{R. Peri\'c$^{1}$, N. Hoffmann$^{2,3}$, A. Chabchoub$^{3,4}$}
\affiliation{$^1$ Institute of Fluid Dynamics and Ship Theory, Hamburg University of Technology, 21073 Hamburg, Germany}
\affiliation{$^2$ Dynamics Group, Hamburg University of Technology, 21073 Hamburg, Germany}
\affiliation{$^3$ Department of Mechanical Engineering, Imperial College London, London SW7 2AZ, United Kingdom}
\affiliation{$^4$ Centre for Ocean Engineering Science and Technology, Swinburne University of Technology, Hawthorn, Victoria 3122, Australia}



\begin{abstract}

The Peregrine breather, today widely regarded as a prototype for spatio-temporally localized rogue waves on the ocean caused by nonlinear focusing, is analyzed by direct numerical simulations based on two-phase Navier-Stokes equations. A finite-volume approach with a volume of fluid method is applied to study the Peregrine breather dynamics up to the initial stages of wave breaking. The comparison of the numerical results with laboratory experiments to validate the numerical approach shows very good agreement and suggests that the chosen method is an effective tool to study modulation instability and breather dynamics in water waves with high accuracy even up to the onset of wave breaking. The numerical results also indicate some previously unnoticed characteristics of the flow fields below the water surface of breathers, which might be of significance for short-term prediction of rogue waves. Recurrent wave breaking is also observed.

\end{abstract}
%


\maketitle


\section{Introduction}

The formation of rogue waves in the oceans is at present intensely debated to be related to modulation instability (MI) \cite{KharifPelinovsky}. This instability was originally discovered in the context of Stokes waves and is thus also often referred to as Benjamin-Feir instability \cite{Bespalov,Lighthill,BF,Zakharov}. Today, it can be discussed most generically within the context of the nonlinear Schr\"odinger equation (NLS) \cite{Newell,Zakharov}, which is the lowest order model for weakly nonlinear dispersive wave envelope dynamics. The NLS is integrable \cite{ZakharovShabat} and has successfully proven to provide suitable initial and boundary conditions to allow observation of soliton dynamics in dispersive and nonlinear media. A special class of exact solutions of the NLS are the so-called breathers on finite background \cite{DystheTrulsen}, which describe strong nonlinear focusing of waves, and therefore rogue wave dynamics. Among the different kinds of breather solutions, there is the Peregrine breather \cite{Peregrine}, which is localized in both time and space. Basically, it describes the nonlinear stage of the MI of Stokes waves for infinite modulation wavelength. Today the Peregrine breather is thus often considered to be the most likely prototype for rogue waves \cite{Shrira,WANDT}. The Peregrine breather itself and related NLS solutions are currently intensely analyzed and studied in several nonlinear dispersive media \cite{KPS,Osborne,OnoratoReport,Zakharov2013}. The recent observations in optics \cite{Kibler,Kibler2}, in water waves \cite{Chabchoub,Chabchoub2,OnratoOMAE,OnoratoPLOSONE,ChabchoubPRE} and in plasma \cite{Bailung} have demonstrated impressively the ability of the NLS to model nonlinear focusing and spatio-temporal localization of wave groups. 

For water waves, a number of studies on breather type dynamics and nonlinear focusing have become available in recent years, and all of them show remarkable correspondence between NLS theory and experimental data for wave states with small steepness, i.e. truly weak nonlinearity. The behaviour of steeper, or more nonlinear waves, and the behaviour of breaking, is far less understood, however. Interestingly, most available studies focus on the surface elevation dynamics only, and it seems that hardly any attention has been paid to the sub-surface flow fields of breather type water waves. In this study, we thus report direct numerical simulations (DNS) of Peregrine breather dynamics on the basis of two-phase incompressible Navier-Stokes equations. A finite volume method (FVM) is applied for discretization and a volume of fluid technique (VOF) for capturing the interface dynamics between water and air. A special focus is put on the dynamic evolution up to initiation of wave breaking, and on the sub-surface flow fields.

\section{Theoretical preliminaries and numerical setup}

The temporal and spatial dynamics of deep-water wave packets can be described by the following form of the NLS \cite{Zakharov,YuenLake},
\begin{equation}
-i\left(\dfrac{\partial A}{\partial t}+c_g\dfrac{\partial A}{\partial x}\right)+\dfrac{\omega_{0}}{8k_0^2}\dfrac{\partial^{2}A}{\partial x^2}+\dfrac{\omega_0k_0^2}{2} |A|^{2}A =0,
\label{NLS}
\end{equation}
with the free surface elevation given by 
\begin{equation}
\eta\left(x,t\right)=\operatorname{Re}\left(A\left(x,t\right)\exp\left[i\left(k_0-\omega_0 t\right)\right]\right).
\label{eta}
\end{equation}
Here, $k_0$ and $\omega_0$ denote the wave number and wave frequency, respectively. For deep-water conditions, the wave-packet $A\left(x,t\right)$ propagates with the group velocity $c_g=\omega_0/2k_0$ being half the phase speed of the waves. The NLS admits an infinite number of pulsating solutions on finite background, which describe the finite amplitude modulation instability dynamics of the Stokes solution $A_S\left(x,t\right)=a_0\exp\left(-i\dfrac{a_0^2k_0^2\omega_0}{2}t\right)$ with an amplitude of $a_0$. The first family of breather solutions that was found are referred to as Akhmediev breathers \cite{Akhmediev1,Akhmediev2}. The solutions are periodic in space and localized in time. The Peregrine breather \cite{Peregrine} arises in the limit of infinite period of the Akhmediev breathers. This particular solution is localized in both space and time. It amplifies the amplitude of the background by a factor of three and is therefore considered to be an appropriate model to describe rogue wave dynamics in water waves as well as other nonlinear dispersive media, such as in nonlinear fiber optics \cite{Kibler} and in plasma \cite{Bailung}. The Peregrine solution of the NLS can be expressed as follows: 
\begin{equation}
A_P\left(x,t\right)=a_0\exp\left(-i\frac{\displaystyle
a_0^2k_0^2\omega_0}{\displaystyle
2}t\right)\left(-1+\frac{\displaystyle4\left(1-ik_0^2a_0^2\omega_0t\right)}{\displaystyle1+\left[2\sqrt{2}k_0^2a_0\left(x-c_gt\right)\right]^2+k_0^4a_0^4\omega_0^2t^2}\right).
\label{Peregrine}
\end{equation}
It is the lowest-order solution of an infinite hierarchy of doubly-localized solutions, derived by nonlinear superposition of several Peregrine solutions \cite{AkhmedievPRE}. The Peregrine breather dynamics has been recently confirmed in water wave experiments \cite{Chabchoub,Chabchoub2}. Subsequently initial conditions for the numerical simulations as well as for the laboratory experiments are determined by evaluating Eq. (\ref{eta}) for a selected position $x^*$, see \cite{Chabchoub3}.

For our simulations we use a finite volume discretization of the Navier-Stokes equations as implemented in the commercial STAR-CCM+ software \cite{vaddi}. The air and the water are assumed to have constant viscosity and density, with the density of air being $ \rho_{\rm air} = 1.2$ kg$\cdot$m$^3$, the density of water being $ \rho_{\rm water} = 1000$ kg$\cdot$m$^3$, the dynamic viscosity of air being $ \mu_{\rm air} = 1.8 \cdot 10^{-5} \ \mathrm{ Pa\cdot s}$ and the dynamic viscosity of water being $ \mu_{\rm water} = 0.001 \ \mathrm{ Pa\cdot s}$. For capturing the interface between air and water, a VOF method is used in this study \cite{VOF}. The distribution of the phases is given by the volume fraction of each phase, with both fractions adding up to one, $\alpha_{\rm air} + \alpha_{\rm water} = 1$. 
Here $ \alpha_{\rm air} $ denotes the volume fraction of gaseous air and $\alpha_{\rm water}$ labels the volume fraction of liquid water. A value of  $\alpha_{\rm air} = 0.5$ in a control volume would usually indicate that the control volume
contains the interface between the phases.

\begin{figure}[H]
\centering
\includegraphics[scale=0.5]{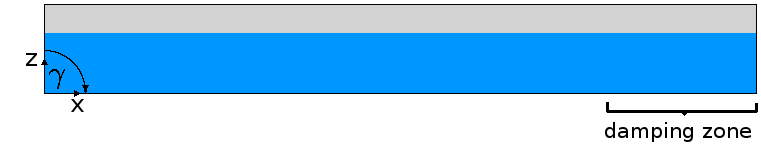}
\caption{The
 configuration of the solution domain. A grid deformation, generating the waves, initiates at $x=0$ and a damping zone extends over 2.5 m on the opposite end.} \label{FIGgeom}
\end{figure}

The solution domain is based on the wave tank used in \cite{Chabchoub2}, see Fig. \ref{FIGgeom} for a schematic sketch. 
The tank is initially filled up to a height of  $ 1\ \mathrm{m} $ with water, while the remaining rest of the tank volume is air at an initial reference pressure of $p_{\mathrm{ref}} = 1013.25 \ \mathrm{hPa}$. 

The origin of the reference coordinate system is located at the bottom left corner of the computational domain. The length in $x$-direction is $ l_{x} = 12.0 \ \mathrm{m} $ and in $z$-direction is $ l_{z} = 1.5 \ \mathrm{m} $. The left wall can rotate around the $y$-axis, characterized by the angle $ \gamma $ relative to the $x$-axis. Thus, the wall moves like a flap type wave maker and can be used to generate waves just as in the experiments \cite{Chabchoub}. The front and back boundaries are considered to be symmetry planes. The left and right tank sides and the tank bottom are modeled as no-slip walls. Towards the end of the experimental tank there is an absorbing beach. Analogously a numerical wave damping zone was introduced into the numerical model, extending over a length of $ 2.5 \ \mathrm{m}$ next to the tank boundary opposite to the moving flap. The damping is implemented by applying a resistance to vertical motions \cite{damp}. For the vertical velocity component $w$ the damping is achieved by adding a source term to the equations of motion, which has the following form:
\begin{equation}
q^{\rm d}_{z} = \rho (f_{1} + f_{2}|w|)\frac{\exp\left(\kappa\right) - 1}{\exp\left(1\right) - 1} w, \quad
\kappa = \left( \frac{x - x_{\rm start}}{x_{\rm end} - x_{\rm start}} \right)^{n_{j}}.
\label{damp}
\end{equation}
Here, $ x $ denotes the wave propagation direction with $ x_{\rm start} $ being the start- and $ x_{\rm end} $ the end-coordinate of the damping zone, respectively. For the present study, the model damping parameter values have been set to  $ f_{1} = 10.0 $, $ f_{2} = 10.0$ and $ n_{j} = 2.0$. From the subsequent results it will be seen that this implementation functions very well. Nevertheless, both in experiments and in numerical analysis we have made sure that there are no spurious effects of reflections coming into play for the results that we discuss below.

The solution domain was discretized with a rectilinear grid. The grid has 200 cells per carrier wavelength $\lambda_{0}$ and 16 cells per carrier wave amplitude $a_{0}$. It is gradually coarsened with increasing distance from the water surface, and the overall number of cells could be substantially reduced in comparison to a grid with constant cell volume, while yielding a comparable discretization error. A number of calculations have been performed to ensure satisfactory convergence of the discretization, and finally a mesh with $685,325$ cells has been selected for the results presented.

\begin{figure}[H]
\begin{center}
\includegraphics[scale=0.25]{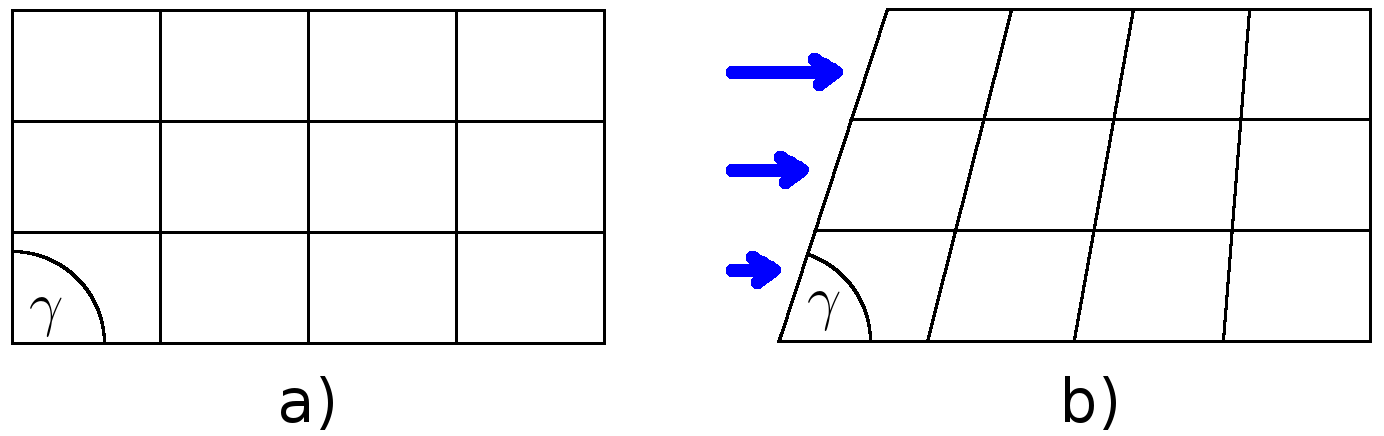}
\end{center}
\caption{Schematical visualization of the morphing process: a) initial grid, b) morphed grid} \label{FIGmorph}
\end{figure}

To generate waves, experimentally \cite{Chabchoub} it has turned out effective to harmonically move the flap with an amplitude proportional to the desired temporal surface elevation $\eta(x^*,t)$ of the water, which in turn is given from the NLS solution. Subsequently, the derivative of $\eta(x,t)$ with respect to time $t$ is calculated numerically. To avoid causing disturbances by starting the flap movement with full amplitude, a linear fade-in for the first $ 4\ \mathrm{s} $ is applied. To realize the flap motion in the computational domain, we used moving control volumes, i.e. moving computational grid cells: the grid was deformed such that the movement of the left domain boundary corresponds to the flap movement. For that purpose, the STAR-CCM+ morpher tool was used. The morpher stretches or shrinks all grid cells in $x$-direction proportional to the distance between the flap wall and the fixed opposite wall, see Fig. \ref{FIGmorph}.  
The advantage of this approach is, that the number of grid cells remains constant. Moreover, the grid retains its quality throughout the morphing process since the flap movement, and thus also the resulting grid deformation, is rather small.

No turbulence modeling was applied to the above equations. Apart from the slight breaking of the rogue wave, the flow is mostly laminar. Thus, the temporal and spatial fluctuations can be resolved with an acceptable computational effort. Turbulence effects, as they might arise due to more severe wave breaking, or due to turbulent boundary layers at the walls of the computational domain, are not considered here.

The STAR-CCM+ Implicit Unsteady solver was applied to carry out the transient computations. To ensure numerical stability in time-marching, a good guideline is that a fluid particle should not cross more than half a computational cell per time step, i.e. with the Courant number $C$, the stability condition is $ C = u_{i} \Delta t/\Delta x_{i} < 0.5$, where $u_{i}$ is the velocity component in $x_{i}$-direction, $\Delta x_{i}$ labels the minimum cell size in $x_{i}$-direction and $\Delta t$ denotes the time-step size. Airy-wave theory gives a reasonable analytical estimate for the maximum velocity components inside a wave as $u_{x,\mathrm{max}} = u_{z,\mathrm{max}} = \omega_{0} a_{0} \exp\left(a_{0}k_{0}\right) \label{Airy_u_max}$. With $ \Delta t_{3} = 0.0005\ \mathrm{s} $, the maximum Courant numbers within a carrier wave result in $C_{x,\mathrm{max}} = 0.025$ in x-direction and  $C_{z,\mathrm{max}} = 0.1$ in z-direction. When the breather waves reach their maximum amplitude of about $3a_{0}$, stability is then still ensured with $C_{z, \mathrm{max}} \approx 0.486 < 0.5$. 

The temporal discretization itself was of second-order, assuming quadratic variation of variables in time. First-order schemes would create significantly more numerical diffusion, which damps the traveling waves inhibiting the formation of strong localizations, as some tests have shown. For each time step, one iteration consists of solving the governing equations for the velocity components, the pressure-correction equation using the SIMPLE method for collocated grids to obtain the pressure values, and the transport equation for the volume fraction. Each equation is linearized, i.e. all other variables except the one for which the equation is solved are assumed as known and taken from previous iteration. $20$ inner iterations were performed per time step, using under-relaxation with factors of  $0.3$ for pressure and $0.95$ for all other variables. The linear equations were solved with an algebraic multigrid method using Gauss-Seidl relaxation schemes, V-cycles for pressure and flexible cycles for velocity calculation. No adjustments to the other default parameters in the software were necessary.

All computations were performed on 16 cores of a Linux cluster and took approximately four weeks to complete. The considerable computational effort resulted primarily from the small time step size, the large amount of inner iterations per time step which were necessary to obtain accurate results, and the comparatively large simulation time of $ 60\ \mathrm{s} $ needed to capture build up and decay of the breather wave packet.

\section{Results of numerical simulations and laboratory experiments}

In the present study, the wavelength of the carrier was set to $\lambda_{0}:=\dfrac{2\pi}{k_0} = 0.5$ m and the background amplitude to $a_0 =0.01$ m. Therefore, the steepness $\varepsilon_0:=a_0k_0=0.1257$, a value which is slightly beyond the breaking threshold $\varepsilon_b=0.12$, see e.g. \cite{Slunyaev}, and the group velocity is $c_g=0.44$ ms$^{-1}$. The initial conditions, applied to the numerical as well as the experimental wave-maker flap, are determined by evaluating Eq. (\ref{eta}) at $x^*=-6$ m. With this, deep-water conditions are satisfied and the applicability of the focusing NLS (\ref{NLS}) is justified. 

\begin{figure}[H]
\centering
\includegraphics[scale=0.087]{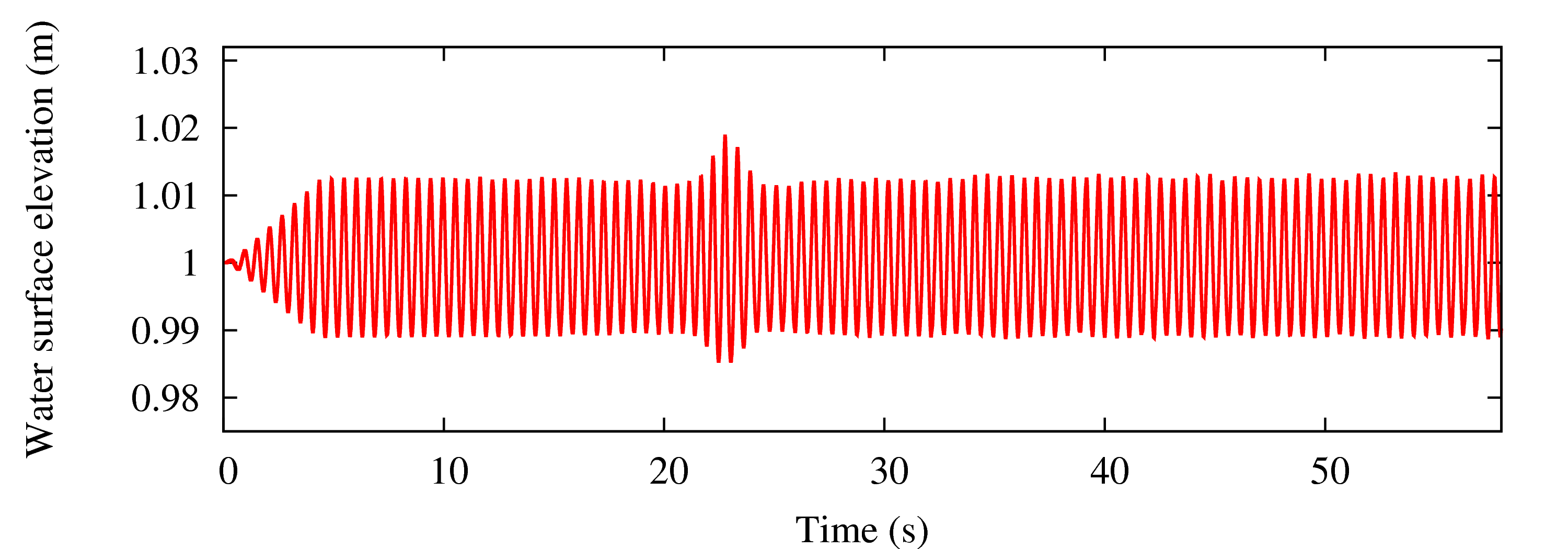}
\includegraphics[scale=0.087]{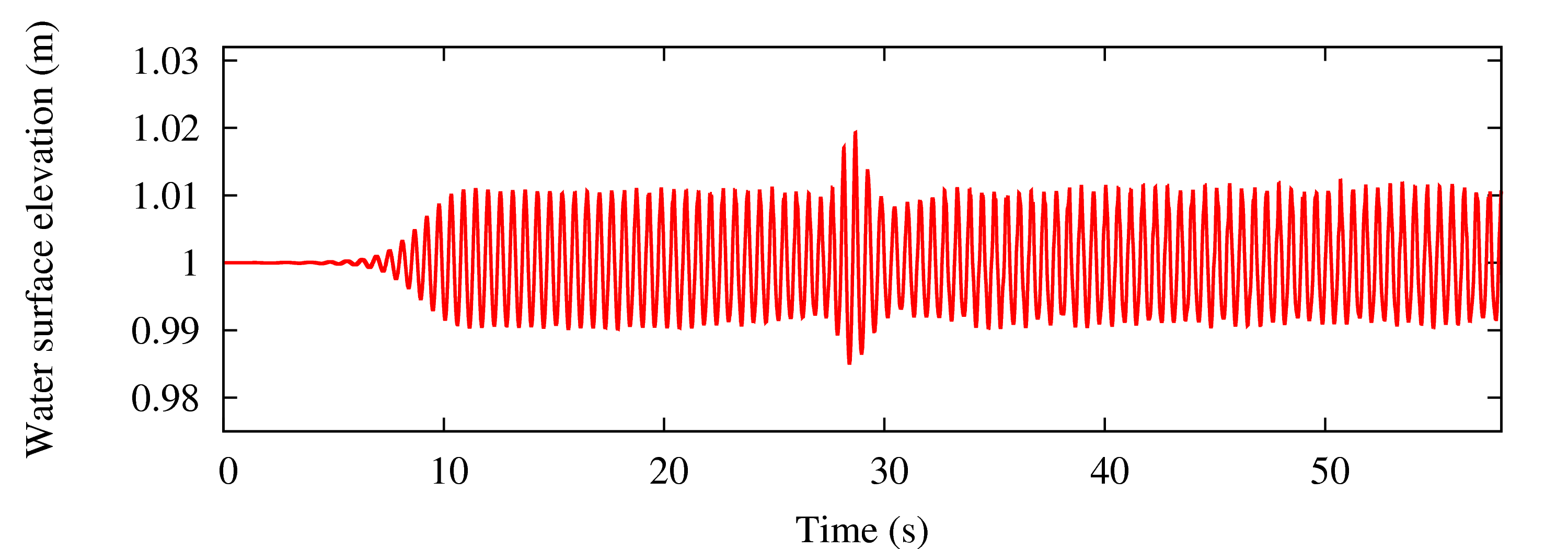}
\includegraphics[scale=0.087]{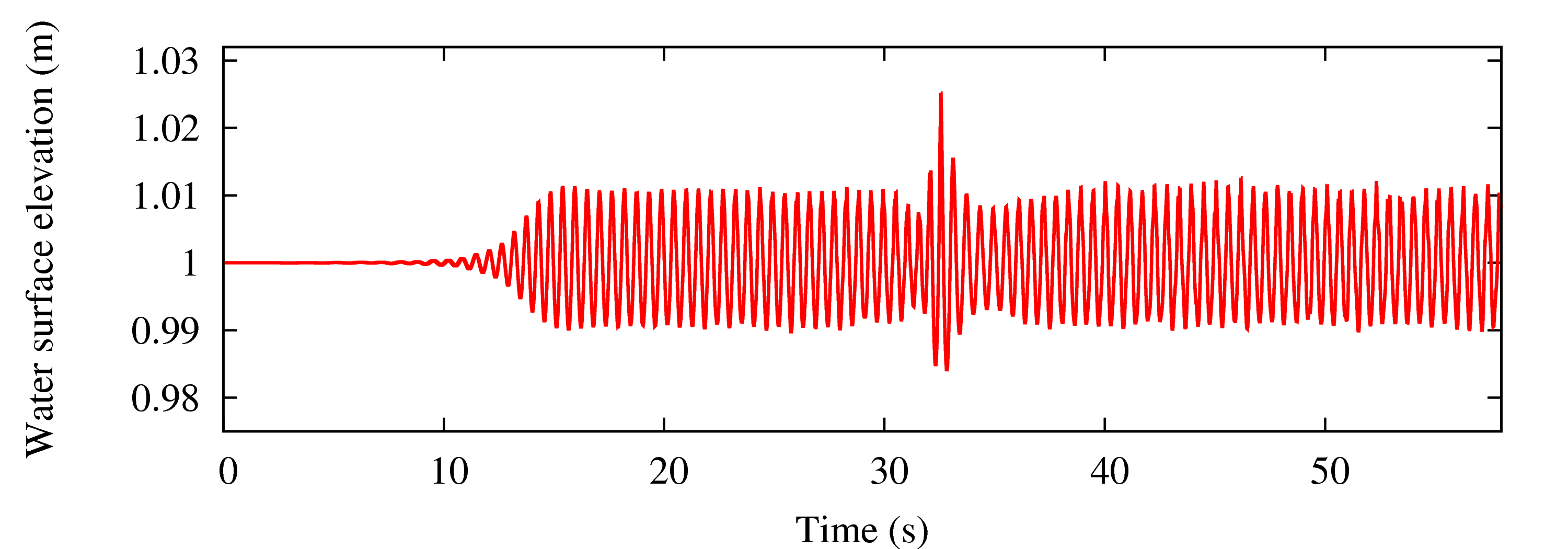}
\includegraphics[scale=0.087]{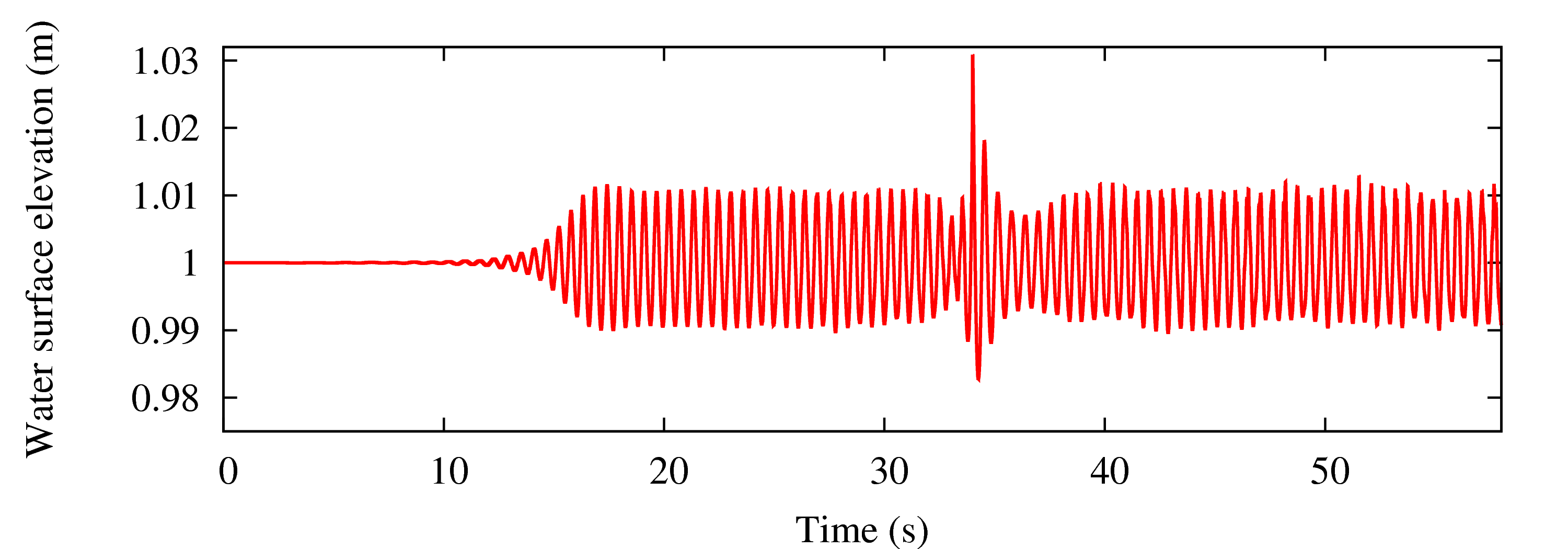}
\includegraphics[scale=0.087]{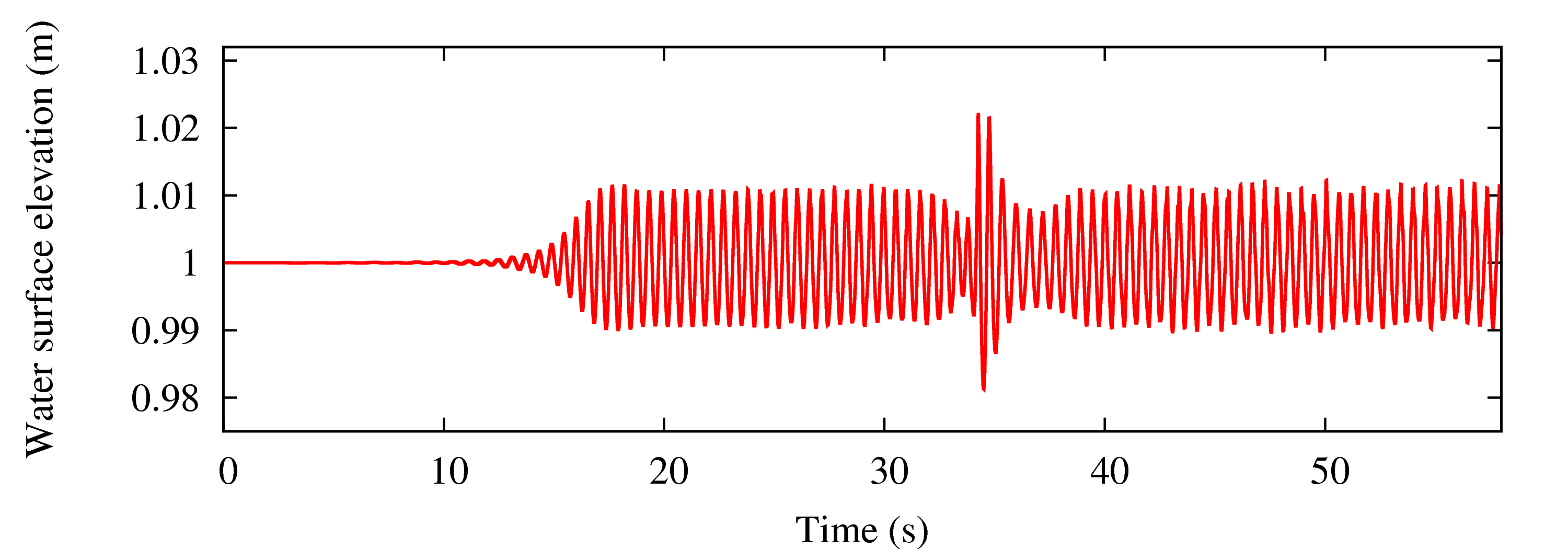}
\caption{Temporal water surface elevation, resulting from the simulation at (from top to bottom) $ x = 0.2\ \mathrm{m} $, $ x = 3.0\ \mathrm{m} $,  $ x = 5.0\ \mathrm{m} $, $ x = 5.8\ \mathrm{m} $ and  $ x = 6.0\ \mathrm{m} $, respectively.} \label{FIGmeas2}
\end{figure}

Fig. \ref{FIGmeas2} shows the results of the numerical simulations of the surface elevation dynamics for the chosen parameters and initial conditions at several positions along the flume.  
As expected from NLS theory, the initially small modulation grows, while evolving in the numerical tank and as previously reported \cite{Chabchoub,Chabchoub2}. Due to the high steepness value selected, and in contrast to the prediction from the NLS, the wave profiles show a marked asymmetry with respect to the vertical axis. In addition, as expected, the amplified waves break before reaching the amplitude amplification of three at a distance of 6 m from flap (which would have been to be expected from NLS theory), since the local steepness there would have exceeded the critical steepness for initiation of breaking, $\varepsilon_0>\varepsilon_b=0.12$. In fact, the breaking occurs at about 5.8 m from the flap, which can be seen from the simulation results both by a near vertical surface there, and by vortical motion in the water.

\begin{figure}[H]
\begin{center}
\includegraphics[scale=0.087]{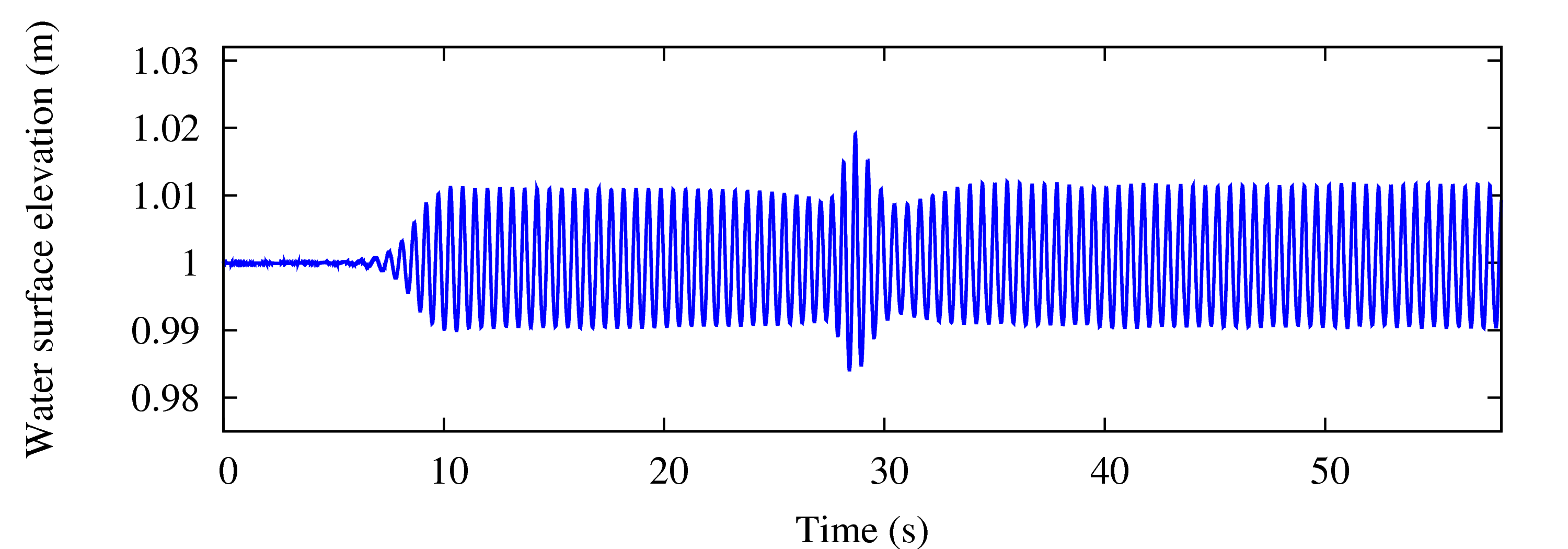}
\includegraphics[scale=0.087]{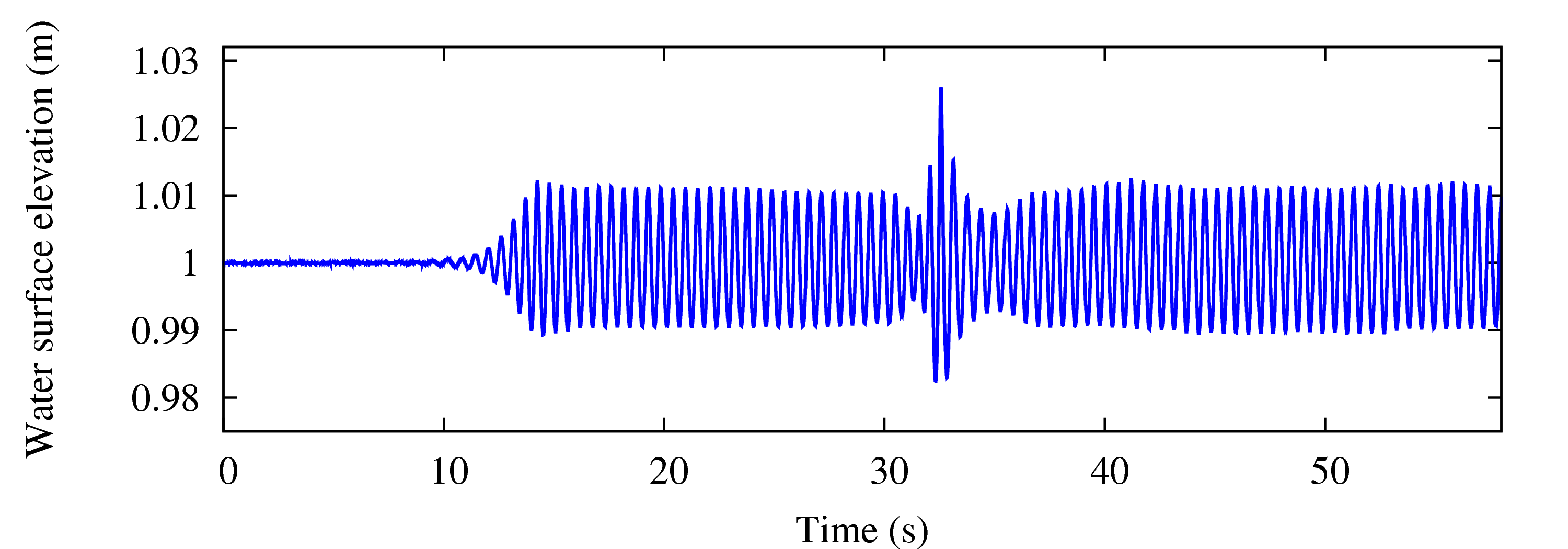}\\
\includegraphics[scale=0.087]{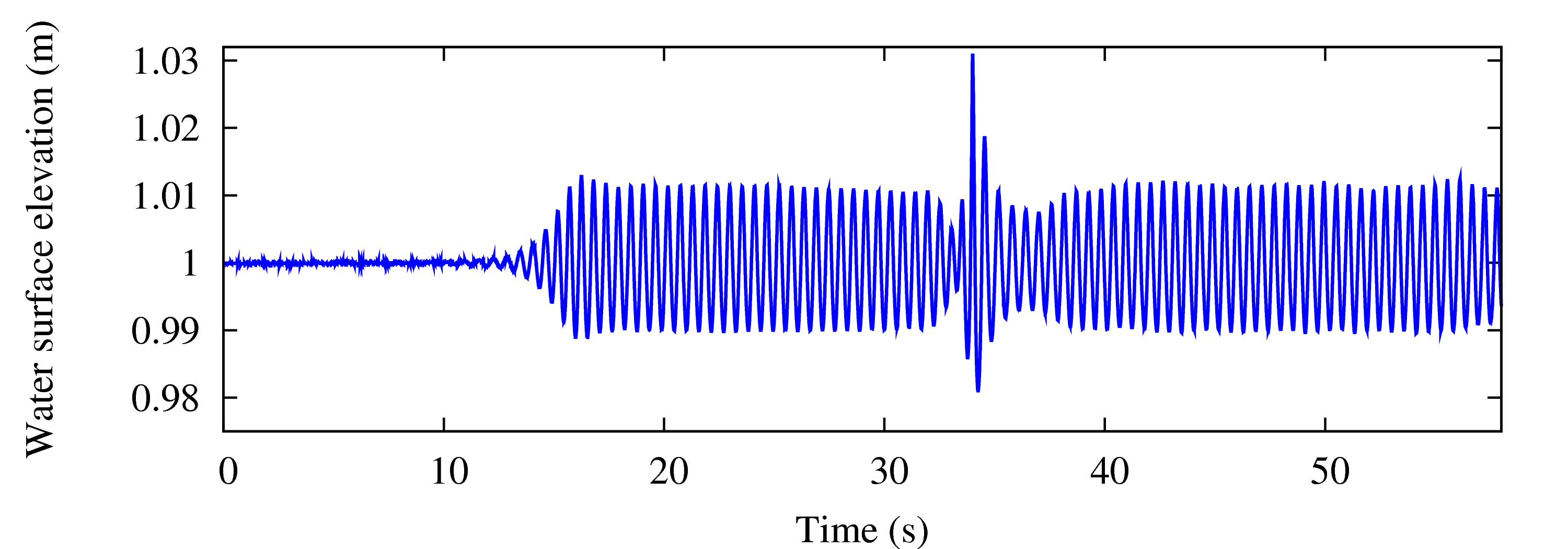}
\includegraphics[scale=0.087]{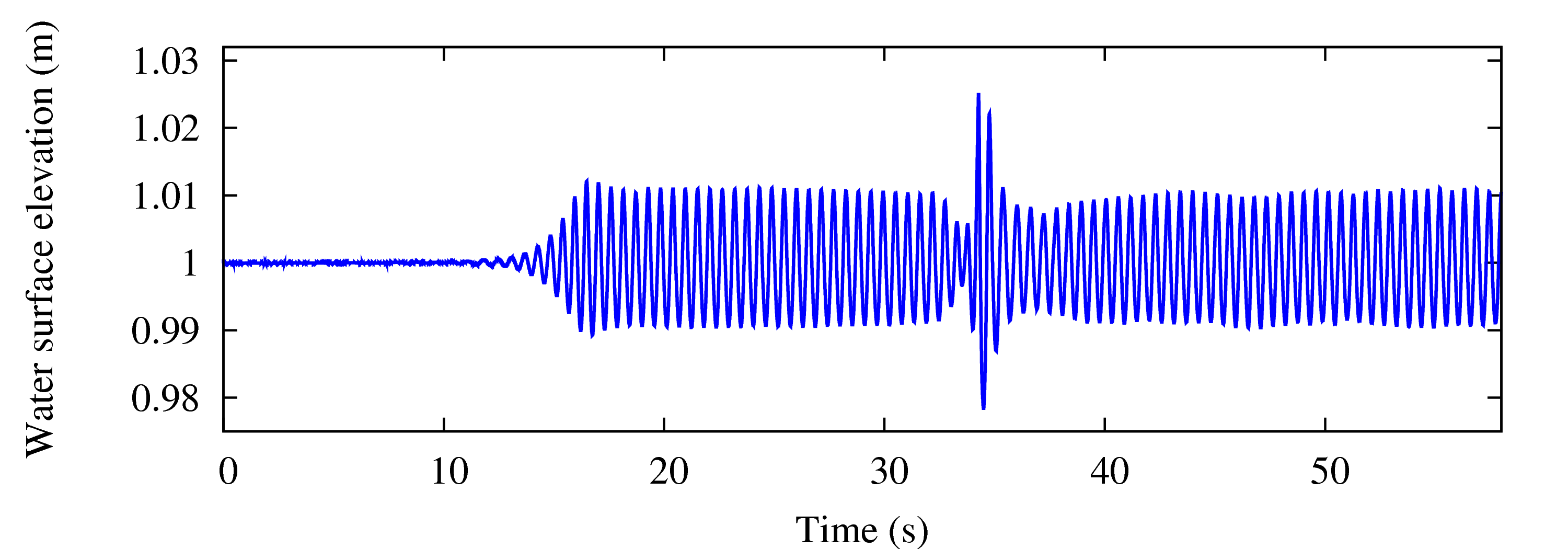}
\end{center}
\caption{Water surface elevation over time from the experiment measured at (from top to bottom) $ x = 3.0 \mathrm{m} $, $ x = 5.0 \mathrm{m} $, $ x = 5.8 \mathrm{m} $ and $ x = 6.0 \mathrm{m} $} \label{FIGexp}
\end{figure}

In
 order to validate the numerical simulations, experiments have been conducted, similar as described in \cite{Chabchoub2}, but with the experimental set-up tuned to fit the present numerical configuration. That is, the background parameters, the initial conditions, the water depth as well as the flap motion have been chosen in correspondence. First experiments have been conducted without cleaning the always somewhat dirty surface of the water. For the initial conditions of the numerical simulations, breaking was then observed at about 6.2 m from the paddle. After cleaning the surface by filtering the water, the breaking position of the maximal wave in the packet was observed 40 cm earlier, i.e. at about 5.8 m from the paddle, which corresponds very well to the numerical simulations. Fig. \ref{FIGexp} shows the experimentally collected temporal measurements of the surface height at the same four positions as already illustrated in Fig. \ref{FIGmeas2} for the simulation.   

\begin{figure}[H]
\begin{center}
\includegraphics[scale=0.087]{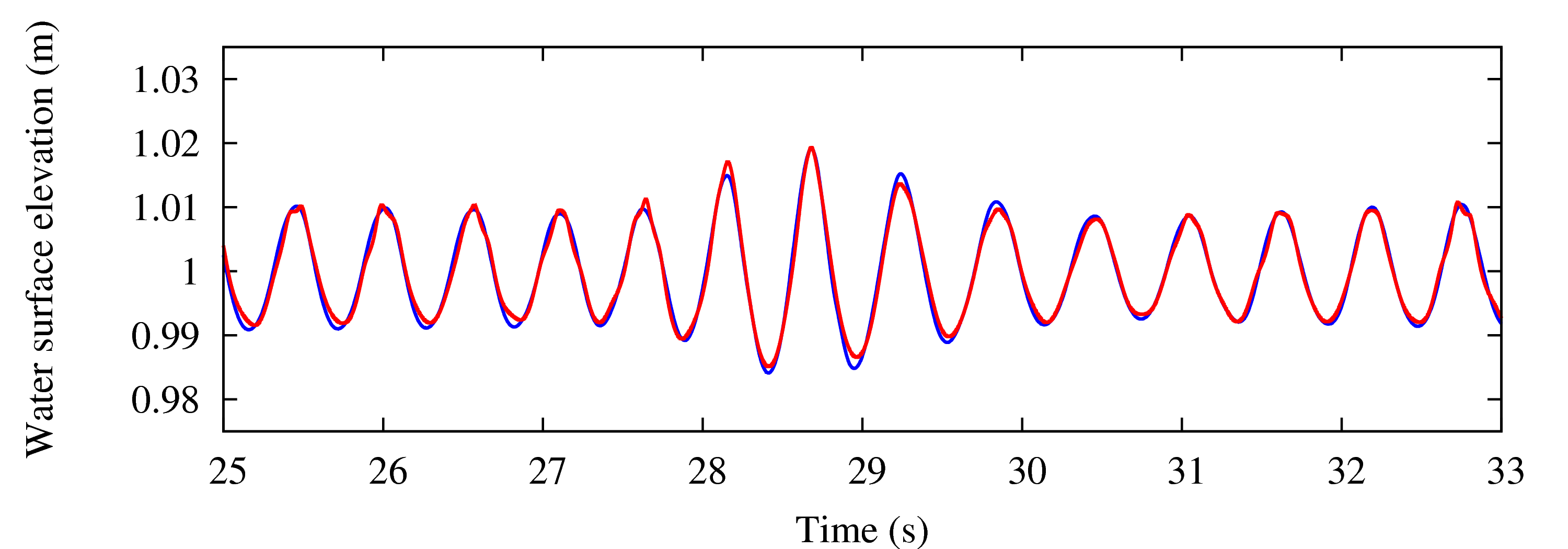}\\
\includegraphics[scale=0.087]{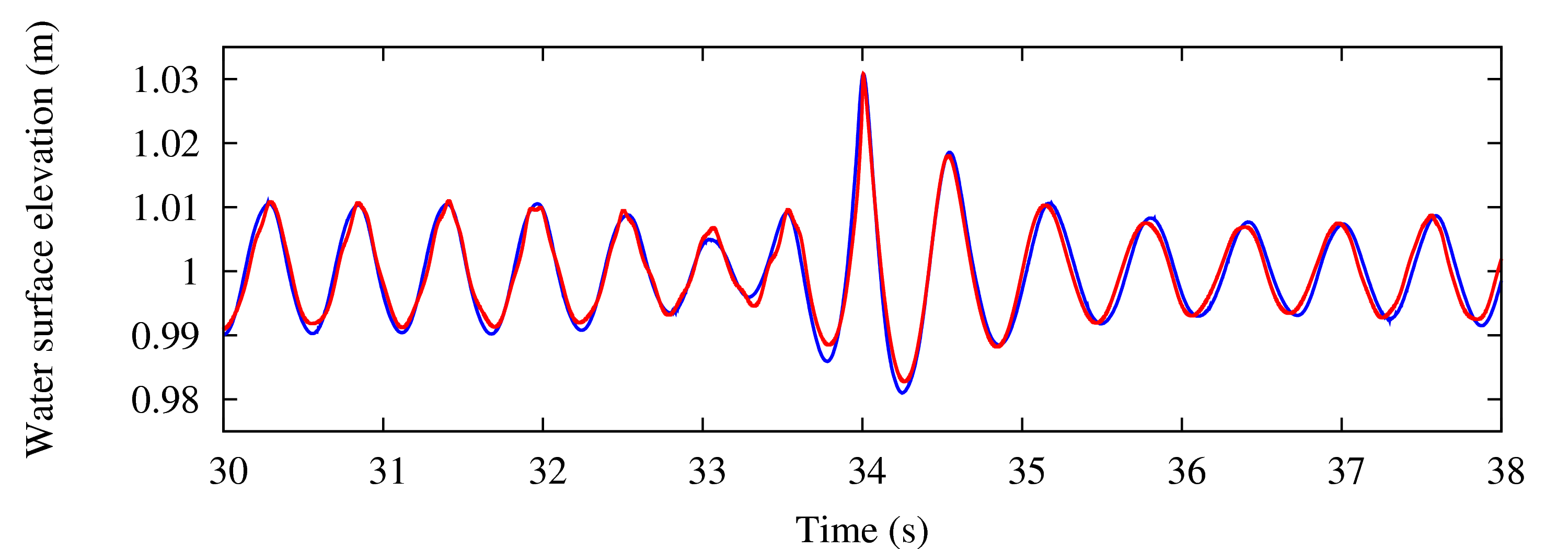}
\end{center}
\caption{Water surface elevation over time at $ x = 3.0 \mathrm{m} $ (top) and at $ x = 5.8 \mathrm{m} $ from experiment (blue lines) and simulation (red lines)} \label{FIGvgl3m}
\end{figure}

As can be noticed, the experimental measurements agree very well with the numerical simulations. In order to demonstrate the excellent agreement in more detail, we superpose the numerical and measured data collected at $x=3.0$ m and at $x=5.8$ m, respectively, see Fig. \ref{FIGvgl3m}.

In contrast to previous comparisons of data with predictions from NLS theory and basic and extended NLS simulation  \cite{Chabchoub2,Slunyaev,Slunyaev2,Dysthe}, here simulation and experiment matches remarkably well and seemingly in detail. 

\begin{figure}[H]
\begin{center}
\includegraphics[scale=0.12]{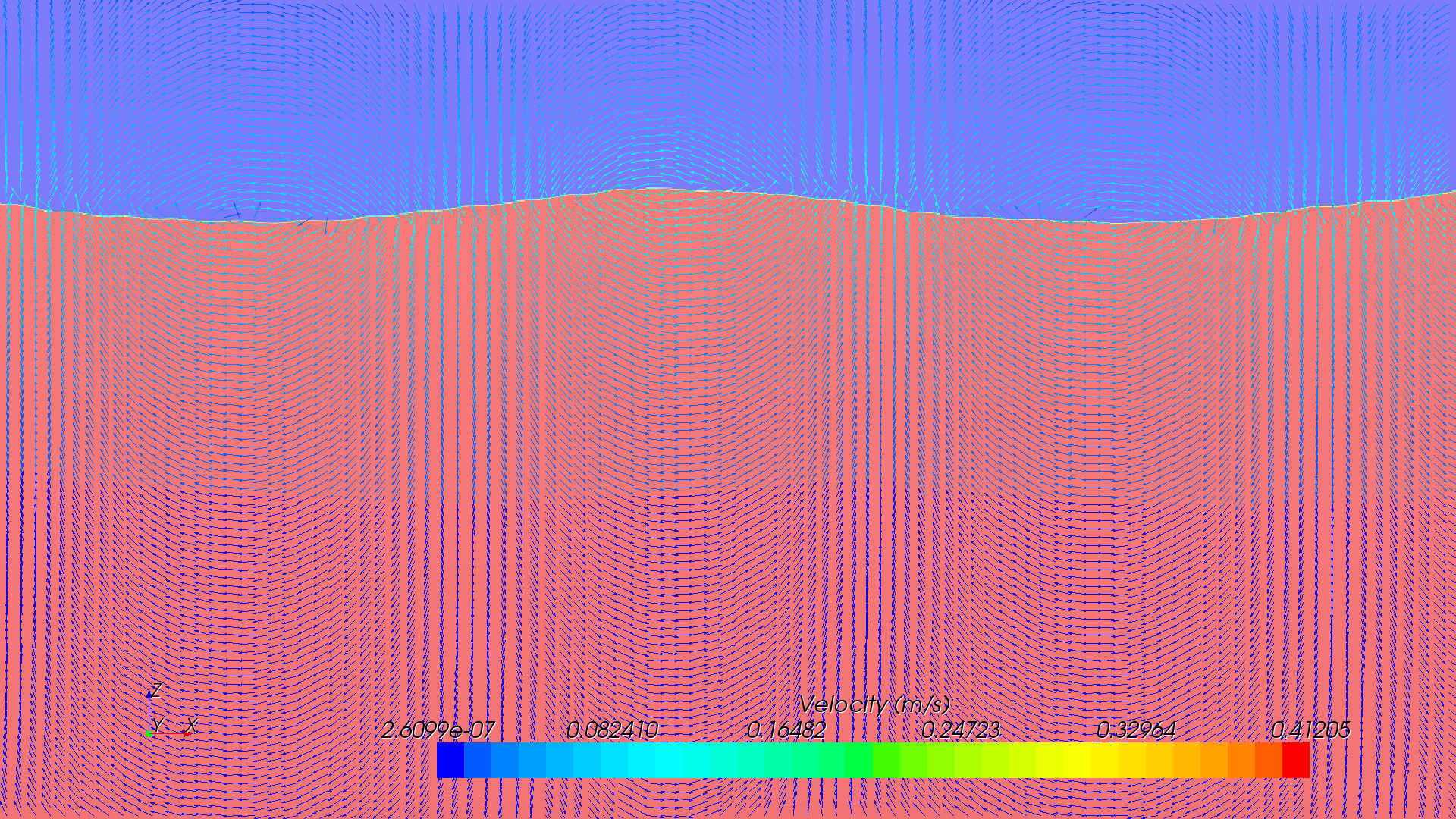}
\includegraphics[scale=0.255]{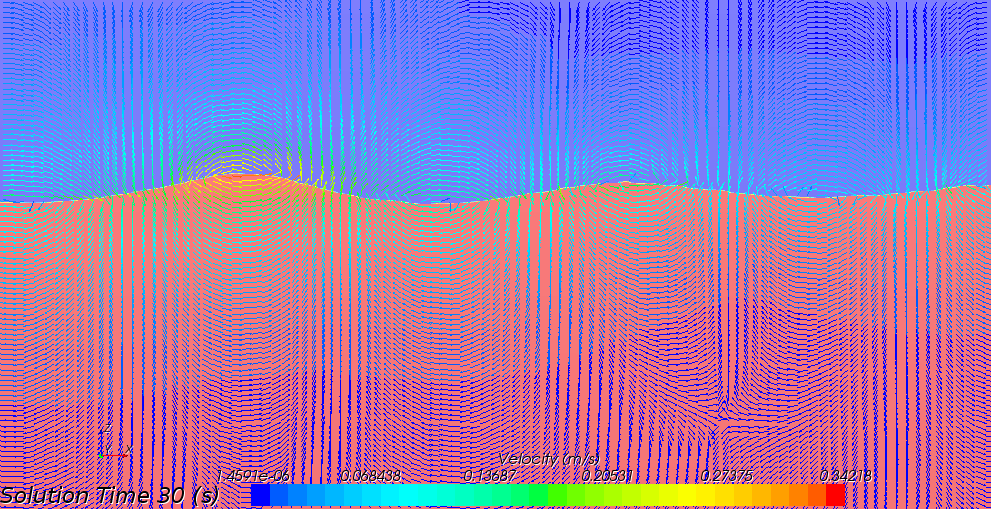}
\includegraphics[scale=0.12]{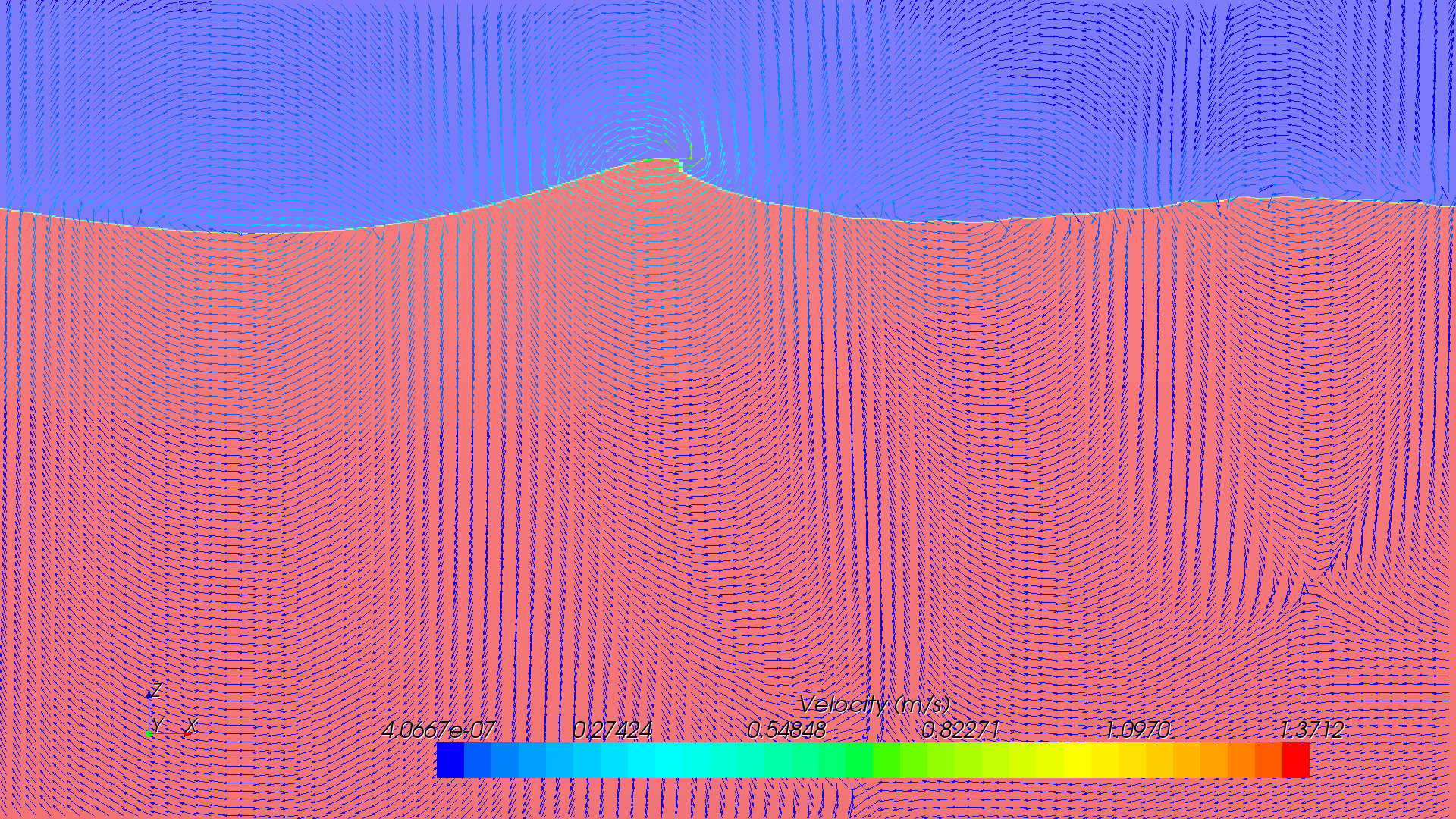}
\end{center}
\caption{Velocity vector fields of: the unmodulated carrier wave (top), the Peregrine modulated Stokes waves at $ t =30.0\ \mathrm{s} $ which correspond to the flume position $ x=3.6\ \mathrm{m} $ (middle) and the modulated waves, just before breaking at $ t = 33.9\ \mathrm{s} $, which correspond to the flume position $ x = 5.8\ \mathrm{m} $ (bottom). The vortical motion near the wave crest indicates the initial stage of breaking.} \label{FIGvv3}
\end{figure}

Another advantage of the present Navier-Stokes based two phase simulation is the possibility of capturing the actual time dependent flow fields of water and air, related to surface motion of the waves. Some results are shown in Fig. \ref{FIGvv3}. 

First one may note from the simulation results at $t=33.9$ s that a small part of the water surface has turned almost vertical near the maximum wave, which is a first indication for the onset of breaking. Second, the flow field around the highest wave crest seems to have become vortical, which is also known as an indication for wave breaking. 

Another interesting result is the behaviour of the sub-surface velocity field when the breather grows. The velocity vector field below the amplified waves becomes asymmetrically distorted, whereas under the undisturbed carrier wave there is a symmetrical velocity field. This asymmetry can already be observed at an early stage of the time evolution of the Peregrine-modulation and seems to be kind of the footprint of the modulation instability on the flow field. It seems tempting to hypothesize that this fundamental property might be used to characterize modulationally unstable waves, or even rogue waves, in the ocean at early stages of their development.

Also an interesting property of water wave breaking in wave packets, here in the course of modulation instability, arises from the difference between the values of group and phase velocities: recurrent breaking \cite{Iafrati}. Once first breaking in a wave packet occurs due to high local wave steepness, it is obvious that the successively amplified waves will be also amplified beyond the local breaking threshold. For the present simulations this recurrence is illustrated in a movie, which can be found in the supplemental material. Recurrent breaking suggests that oceanic rogue waves might be even more dangerous as previously expected, since one unstable wave packet would generate several very steep breaking wave crests, rather than just one single and quickly disappearing breaker. Obviously, the impact of such kind of multiply breaking wave trains on ships or structures could be much more severe when compared to the impact of non-breaking waves, having nearly the same local steepness. 

\section{Conclusion}

We have shown that rogue wave dynamics, related to modulation instability, and NLS breather solutions, can be simulated in numerical computations solving the two-phase Navier-Stokes equations. In particular, we applied the same Peregrine-breather initial condition to the flap of the numerical as well as the laboratory wave flume, respectively. The parameters of the carrier have been chosen in order to consider the breaking in the modulated wave train within the propagation distance in the tank. The numerical simulations show excellent agreement with the laboratory experiments. In addition, the simulations show interesting characteristics of the velocity vector fields of the modulated and unstable nonlinear waves, suggesting a novel criterion for early stage identification and prediction of nonlinear and unstable waves in narrow-banded sea state conditions. Furthermore, it has been shown that the simulations also capture the recurrent breaking phenomenon, which is related to the modulation instability and to the difference between the group and phase velocity values of the waves. The main disadvantage of the presented numerical method is the considerable computational effort. Nevertheless, the approach has proven to provide a powerful tool for analysis and visualization.

\end{document}